\documentclass[a4paper,11pt]{article}
\usepackage[dvipdfmx]{graphicx}
\usepackage{amsmath}        
\usepackage{amssymb}        
\usepackage{cite}
\usepackage{cancel}
\usepackage{fancybox}
\usepackage[dvips]{color}
\usepackage{url}
\usepackage{ulem}
\usepackage{pifont}
\usepackage{slashbox}
\usepackage{colortbl}
\usepackage{multicol}
\usepackage{lscape}
\usepackage{colortbl}
\usepackage{multicol}
\usepackage[all,knot]{xy}
\usepackage{longtable}

\makeatletter
 
 \@addtoreset{equation}{section}
\makeatother

\begin{document}

\title{String-Inspired Special Grand Unification}

\author{Naoki Yamatsu
\footnote{Electronic address: yamatsu@cc.kyoto-su.ac.jp}
\\
{\it\small Maskawa Institute for Science and Culture, 
Kyoto Sangyo University, Kyoto 603-8555, Japan}
}
\date{\today}

\maketitle

\begin{abstract}
We discuss a grand unified theory (GUT) based on an $SO(32)$ GUT gauge
group broken to its subgroups including a special subgroup.
In the $SO(32)$ GUT on six-dimensional (6D) orbifold space
$M^4\times T^2/\mathbb{Z}_2$, one generation of the SM fermions can be
embedded into a 6D bulk Weyl fermion in the $SO(32)$ vector
representation. We show that for a three generation model, all the 6D
and 4D gauge anomalies in the bulk and on the fixed points are canceled
out without exotic chiral fermions at low energies. 
\end{abstract}

\section{Introduction}

Grand unification \cite{Georgi:1974sy}
is one of the most attractive idea to construct unified theories beyond
the standard model (SM).
It is known in e.g., Refs.~\cite{Slansky:1981yr,Yamatsu:2015gut} that
the candidates for grand unified theory (GUT) gauge
groups in four-dimensional (4D) theories are only $SU(n) (n\geq 5)$, 
$SO(4n+2) (n\geq 2)$, and $E_6$ because of rank and type of 
representations, while
the candidates for GUT gauge groups in higher dimensional theories are
$SU(n) (n\geq 5)$, $SO(n) (n\geq 9)$, $USp(2n) (n\geq 4)$, 
$E_n (n=6,7,8)$, and $F_4$.
Many GUTs have been already discussed in e.g.,
Refs.~\cite{Georgi:1974sy,Inoue:1977qd,Fritzsch:1974nn,Ida:1980ea,Fujimoto:1981bv,Gursey:1975ki,Kawamura:2013rj,Kojima:2017qbt}
for 4D GUTs;
Refs.~\cite{Kojima:2011ad,Kojima:2016fvv,Burdman:2002se,Lim:2007jv,Kim:2002im,Fukuyama:2008pw,Hosotani:2015hoa,Hosotani:2015wmb,Yamatsu:2015rge,Furui:2016owe,Hosotani:2016njs}
for 5D GUTs;
Ref.~\cite{Hosotani:2017ghg}
for 6D GUTs.

Recently, a new-type GUT has been proposed by the author in
Ref.~\cite{Yamatsu:2017sgu}. In usual GUTs, GUT gauge groups are 
broken to their {\it regular} subgroups; e.g., 
$E_6\supset SO(10)\times U(1)\supset SU(5)\times U(1)\times U(1)
\supset G_{\rm SM}\times U(1)\times U(1)$.
In the new GUT called a {\it special GUT}, GUT gauge groups are broken 
down to {\it special subgroups}.
(For Lie groups and their regular and special subgroups, see e.g.,
Refs.~\cite{Dynkin:1957ek,Dynkin:1957um,Cahn:1985wk,Yamatsu:2015gut}.)

In Ref.~\cite{Yamatsu:2017sgu}, the author
proposed an $SU(16)$ special GUT whose gauge group $SU(16)$ is broken to
a special subgroup $SO(10)$.  
The results are summarized as follows. 
In a 4D $SU(16)$ special GUT, one generation of quarks and leptons can
be embedded into a 4D $SU(16)$ ${\bf 16}$ Weyl fermion;
the 4D $SU(16)$ gauge anomaly restricts the minimal number of
generations. Unfortunately, the minimal number is 12 in 4D framework.
In a 6D $SU(16)$ special GUT on 6D orbifold space 
$M^4\times T^2/\mathbb{Z}_2$,
one generation of quarks and leptons can
be embedded into a 6D $SU(16)$ ${\bf 16}$ Weyl fermion;
the 6D $SU(16)$ gauge anomaly and the 4D $SU(16)$ gauge anomaly on the
fixed points restricts the minimal number of generations;
three generation of quarks and leptons is allowed without 4D exotic
chiral fermions.

Superstring theory \cite{Polchinski1998a,Polchinski1998b}
has been considered as a candidate for unified
theory to describe all the interaction including gravity.
There are a lot of attempts to construct the SM from string theories.
One of its trials is based on $E_8\times E_8$ and $SO(32)$ heterotic
string theories \cite{Green1984,Green:1984ed,Derendinger:1985kk,Giedt:2003an,Choi:2004wn,Blumenhagen:2005pm,Nilles:2006np,Ito:2010df,Ito:2011ng,Abe:2015mua,Abe:2015xua,Abe:2016eyh}. 
Usually, the $E_8\times E_8$ heterotic string model 
building is much more popular than $SO(32)$ one. One of the biggest
reason seems that when we only consider regular embeddings,
for the branching rules of $SO(32)\supset SO(10)(\times U(1)^{11})$, the
$SO(32)$ vector and adjoint representations ${\bf 32}$ and ${\bf 496}$
do not contain $SO(10)$ spinor representations ${\bf 16}$ and 
${\bf \overline{16}}$, while for the branching rules of 
$E_8\supset SO(10)(\times U(1)^{3})$, the $E_8$ adjoint representation
${\bf 248}$ contain $SO(10)$ spinor representations ${\bf 16}$ and 
${\bf \overline{16}}$.
However, for a special embedding, on the other hand,
the branching rules of 
$SO(32)(\supset SU(16)\times U(1)_Z)\supset SO(10)\times U(1)_Z$ for 
$SO(32)$ vector and adjoint representations ${\bf 32}$ and ${\bf 496}$
are given by
\begin{align}
{\bf 32}=&
({\bf 16})(2)
\oplus({\bf \overline{16}})(-2),
\label{branching-rule-32-SO10}\\
{\bf 496}=&
({\bf 210})(0)
\oplus({\bf 45})(0)
\oplus({\bf 120})(4)
\oplus({\bf 120})(-4)
\oplus({\bf 1})(0),
\label{branching-rule-496-SO10}
\end{align}
where we follow the convention of their $U(1)$ normalization in
Ref.~\cite{Yamatsu:2015gut}. 
Obviously, an $SO(32)$ vector
representation is decomposed into $SO(10)$ spinor representations, and
an $SO(32)$ adjoint representation contains an $SO(10)$ bi-spinor
representation ${\bf 210}$.
When we take into account the special embedding $SU(16)\supset SO(10)$,
$SO(32)$ gauge theories contain $SO(10)$ spinors, easily.
In the following discussion, we will not consider how to realize models
from string theories.

There are several good features of special GUTs pointed out in
Ref.~\cite{Yamatsu:2017sgu}. First, almost all unnecessary $U(1)$s 
can be eliminated; e.g., $SO(32)\supset G_{\rm SM}\times U(1)^{12}$ 
by using only regular embeddings, 
while $SO(32)(\supset SU(16)\times U(1)_Z\supset SO(10)\times U(1)_Z)
\supset G_{\rm SM}\times U(1)^2$ by using regular and special embeddings.
Second, by using only regular embeddings, 
the SM fermions cannot be embedded into an $SO(32)$ vector
representation ${\bf 32}$, 
while by using regular and special embeddings 
$SO(32)(\supset SU(16)\times U(1)_Z\supset SO(10)\times U(1)_Z)
\supset G_{\rm SM}\times U(1)^2$,
the SM fermions can be embedded into an $SO(32)$ vector representation.

It is known in e.g., Refs.~\cite{Slansky:1981yr,Yamatsu:2015gut} that
any 4D $SO(32)$ gauge theory is a vectorlike theory since an $SO(32)$
group has only real representations. To realize the SM,
i.e., a 4D chiral gauge theory, we take orbifold space construction
\cite{Kawamura:1999nj,Kawamura:2000ev}.
It allows us to realize 4D Weyl fermions from 5D Dirac fermions, 6D Weyl
fermions, etc.
In the 6D $SU(16)$ special GUT \cite{Yamatsu:2017sgu}, 
the nonvanishing VEV of a 5D $SU(16)$ ${\bf 5440}$ brane scalar is
responsible to break the $SU(16)$ GUT gauge group to its special
subgroup $SO(10)$ via the Higgs mechanism
\cite{Higgs:1964ia,Higgs:1964pj}.
For $SO(32)$ special GUTs, the $SO(32)$ GUT gauge group can be broken to 
$SO(10)$ by using the nonvanishing VEV of a scalar in an appropriate
representation of $SO(32)$; the lowest dimensional representation is
{\bf 86768}. 
(The spontaneous symmetry breaking of $SU(n)$ to its special subgroups 
has been discussed in e.g., Refs.~\cite{Li:1973mq,Meljanac:1982rc}.)

In this paper, we will discuss an $SO(32)$ special GUT on 6D orbifold
spacetime $M^4\times T^2/\mathbb{Z}_2$. As in 6D $SU(16)$ special GUTs,
we need to take into account 6D and 4D gauge anomalies. As the same as
the 6D $SU(16)$ gauge anomaly in the 6D $SU(16)$ special GUT
\cite{Yamatsu:2017sgu}, the 6D $SO(32)$ gauge anomaly can be canceled
out by introducing 6D positive and negative Weyl fermions in the same
representation of $SO(32)$ gauge group. Unlike an $SU(16)$ gauge group,
an $SO(32)$ gauge group itself has no 4D gauge anomaly for any fermion
in any representation of $SO(32)$, but there can be 4D gauge anomalies
for its subgroups. We will see it in Sec.~\ref{Sec:Special-GUT} in
detail. 

The main purpose of this paper is to show that 
in a 6D $SO(32)$ special GUT on $M^4\times T^2/\mathbb{Z}_2$
we can realize three generations of the 4D SM Weyl fermions 
from six 6D $SO(32)$ ${\bf 32}$ bulk Weyl fermions without 4D exotic
chiral fermions at low energies, and without any 6D and 4D gauge
anomaly.

This paper is organized as follows. In Sec.~\ref{Sec:basics}, before we
discuss a special GUT based on an $SO(32)$ gauge group, we quickly
review basic properties of $SO(32)$ and its subgroups
shown in Ref.~\cite{Yamatsu:2015gut}.
In Sec.~\ref{Sec:Special-GUT}, 
we construct a 6D $SO(32)$ special GUT on $M^4\times T^2/\mathbb{Z}_2$.
Section~\ref{Sec:Summary-discussion} is devoted to a summary and
discussion.

\section{Basics for $SO(32)$ and its subgroups}
\label{Sec:basics}

First, we check how to embed the SM Weyl fermions into $SO(32)$ vector
multiplets.
For regular and special embeddings
$SO(32)\supset SU(16)\times U(1)_Z\supset SO(10)\times U(1)_Z$,
an $SO(32)$ vector representation ${\bf 32}$ is decomposed into 
$SO(10)$ spinor representations ${\bf 16}$ and ${\bf \overline{16}}$.
Further, the $SO(10)$ spinor representation ${\bf 16}$ is decomposed into 
$G_{\rm SM}\times U(1)_X=SU(3)_C\times SU(2)_L\times U(1)_Y\times U(1)_X$
representations:
\begin{align}
{\bf 16}=&
({\bf 3,2})(-1)(1)
\oplus({\bf \overline{3},1})(-2)(-3)
\oplus({\bf \overline{3},1})(4)(1)\nonumber\\
&\oplus({\bf 1,2})(3)(-3)
\oplus({\bf 1,1})(-6)(1)
\oplus({\bf 1,1})(0)(5).
\end{align}
Since the $SO(32)$ vector representation ${\bf 32}$ is real, a 4D Weyl
fermion in $SO(32)$ ${\bf 32}$ representation includes not only 4D SM
Weyl fermions but also their conjugate fermions. To realize chiral
fermions, we take orbifold symmetry breaking mechanism
\cite{Kawamura:1999nj,Kawamura:2000ev}.
After taking into account orbifold effects, we can regard the zero modes
of an $SO(32)$ ${\bf 32}$ fermion as one generation of the SM fermions
plus a right-handed neutrino. Note that there is no 4D pure $SO(32)$
gauge anomalies of any representation of $SO(32)$ gauge group, while
there can be 4D $SU(16)$ and $U(1)$ anomalies generated by 4D Weyl
fermions in complex representations of $SU(16)$ and $U(1)$,
respectively. Then, after orbifolding, a maximal regular subgroup
$SU(16)\times U(1)_Z$ of $SO(32)$ may be anomalous.
We will discuss how to cancel out 4D pure $SU(16)$, pure $U(1)_Z$, mixed
$SU(16)-SU(16)-U(1)_Z$ and mixed $\mbox{grav.}-\mbox{grav.}-U(1)_Z$ 
generated by 6D bulk fermions in the next section.  

We consider a symmetry breaking pattern from $SO(32)$ to $G_{\rm SM}$. 
One way of achieving it is to use orbifold symmetry breaking boundary
conditions (BCs) and several GUT breaking Higgses. One example is 
to choose orbifold BCs breaking $SO(32)$ to $SU(16)\times U(1)$ and 
to introduce three $SO(32)$ ${\bf 86768}$, ${\bf 496}$, 
${\bf 32}$ scalar fields, where we assume their proper components get
non-vanishing VEVs.  
First, the following orbifold BC for the $SO(32)$ vector representation 
${\bf 32}$ breaks $SO(32)$ to $SU(16)\times U(1)_Z$:
\begin{align}
P_{\bf 32}=
\sigma_2\otimes I_2\otimes I_2\otimes I_2\otimes I_2,
\end{align}
where the projection matrix $P_{\bf 32}$ is proportional to the $U(1)_Z$
generator and satisfies $(P_{\bf 32})^2=I_{32}$. (The matrix form of
$P_{\bf 32}$ depends on basis.)
Next, the non-vanishing VEV of the $SO(32)$ ${\bf 86768}$ scalar
field is responsible for breaking $SO(32)\supset SU(16)\times U(1)_Z$ to
$SO(10)$ or $SO(10)\times U(1)_Z$, where its branching rule of 
$SO(32)\supset SU(16)\times U(1)_Z$ is given by
\begin{align}
{\bf 86768}=&
({\bf 18240})(0)
\oplus({\bf 14144})(0)
\oplus({\bf 5440})(8)
\oplus({\bf \overline{5440}})(-8)
\oplus({\bf 255})(0)
\oplus({\bf 1})(0)
\nonumber\\
&
\oplus({\bf 21504})(4)
\oplus({\bf \overline{21504}})(-4)
\oplus({\bf 120})(4)
\oplus({\bf \overline{120}})(-4).
\label{branching-rule-86768}
\end{align}
$SU(16)$ ${\bf 18240}$ and ${\bf 5440}$ $({\bf \overline{5440}})$
contain singlet under its $SO(10)$ special subgroup. 
Their nonvanishing VEV can break $SU(16)$ to its special subgroup
$SO(10)$ \cite{Yamatsu:2017sgu}, where their $SO(10)$ decompositions are
given in Ref.~\cite{Yamatsu:2015gut} by 
\begin{align}
{\bf 18240}=&
({\bf 8910})
\oplus({\bf 5940})
\oplus({\bf 770})
\oplus({\bf 1050})
\oplus({\bf \overline{1050}})
\oplus({\bf 54})
\oplus2({\bf 210})
\oplus({\bf 45})
\oplus({\bf 1}),
\label{branching-rule-18240}\\
{\bf 5440}=&
{\bf 4125}
\oplus{\bf \overline{1050}}
\oplus{\bf 210}
\oplus{\bf 54}
\oplus{\bf 1}\ \ \
({\bf \overline{5440}}=
{\bf 4125}
\oplus{\bf 1050}
\oplus{\bf 210}
\oplus{\bf 54}
\oplus{\bf 1}).
\label{branching-rule-5440}
\end{align}
The VEV of the $SO(32)$ ${\bf 32}$ scalar, then,  breaks 
$(SO(32)\supset)SO(10)\times U(1)_Z$ to $SU(5)$,  
where its branching rule of 
$SO(32)\supset SU(16)\times U(1)_Z$ is given by
\begin{align}
{\bf 32}=&
({\bf 16})(2)
\oplus({\bf \overline{16}})(-2),
\label{branching-rule-32}
\end{align}
The VEV of the $SO(32)$ ${\bf 496}$ scalar further
reduces $(SO(32)\supset SU(16)\supset SO(10)\supset)SU(5)$ 
to $G_{\rm SM}$, 
where its branching rule of $SO(32)\supset SU(16)\times U(1)_Z$ is given
by
\begin{align}
{\bf 496}=&
({\bf 255})(0)
\oplus({\bf 120})(4)
\oplus({\bf \overline{120}})(-4)
\oplus({\bf 1})(0),
\label{branching-rule-496}
\end{align}
and the $SU(16)$ adjoint representation ${\bf 255}$
is decomposed into $SO(10)$ representations 
\begin{align}
{\bf 255}={\bf 210}\oplus{\bf 45}.
\label{branching-rule-255}
\end{align}
(For further information, see e.g., Ref.~\cite{Yamatsu:2015gut}.)

\section{$SO(32)$ special grand unification}
\label{Sec:Special-GUT}

We start to discuss an $SO(32)$ special GUT on 6D orbifold
spacetime $M^4\times T^2/\mathbb{Z}_2$ with 
the Randall-Sundrum (RS) type metric
\cite{Randall:1999ee,Hosotani:2017ghg}
given by 
\begin{align}
ds^2=e^{-2\sigma(y)}(\eta_{\mu\nu}dx^{\mu}dx^{\nu}+dv^2)+dy^2,
\end{align}
where $y$ is the coordinate of RS warped space,
$v$ is the coordinate of $S^1$,
$\eta_{\mu\nu}=\mbox{diag}(-1,+1,+1,+1)$, 
$\sigma(y)=\sigma(-y)=\sigma(y+2\pi R_5)$,
$\sigma(y)=k|y|$ for $|y|\leq \pi R_5$, and
$v\sim v+2\pi R_6$.
There are four fixed points on $T^2/\mathbb{Z}_2$ 
at $(y_0,v_0)=(0,0)$, $(y_1,v_1)=(\pi R_5,0)$, 
$(y_2,v_2)=(0,\pi R_6)$, and $(y_3,v_3)=(\pi R_5,\pi R_6)$.
For each fixed point, the $\mathbb{Z}_2$ parity reflection 
is described by 
\begin{align}
P_j:\ (x_\mu,y_j+y,v_j+v)\ \to\ (x_\mu,y_j-y,v_j-v),
\end{align}
where $j=0,1,2,3$, and $P_3=P_1P_0P_2=P_2P_0P_1$.
5th and 6th dimensional translation 
$U_5: (x_\mu,y,v)\to(x_\mu,y+2\pi R_5,v)$ 
and $U_6: (x_\mu,y,v)\to(x_\mu,y,v+2\pi R_6)$  
satisfy $U_5=P_1P_0$ and $U_6=P_2P_0$, respectively.

We consider the matter content in the $SO(32)$ special GUT that consists
of a 6D $SO(32)$ bulk gauge boson $A_{M}$;
three 6D $SO(32)$ ${\bf 32}$ positive Weyl fermions with the orbifold BCs
$(\eta_{0},\eta_{1},\eta_{2},\eta_{3})=(-,-,-,-)$ 
$\Psi_{{\bf 32}+}^{(a)}$ $(a=1,2,3)$ and 
three 6D negative one with $(-,+,-,+)$
$\Psi_{{\bf 32}-}^{(b)}$ $(b=1,2,3)$;
5D $SO(32)$ ${\bf 86768}$, ${\bf 496}$ and ${\bf 16}$ brane scalar
bosons at $y=0$ 
$\Phi_{\bf 86768}$, $\Phi_{\bf 496}$, $\Phi_{\bf 32}$;
a 4D $SU(16)\times U(1)$ $({\bf \overline{120}})(0)$ Weyl brane fermion,
twelve 4D $SU(16)\times U(1)$ 
$({\bf {16}})(0)\oplus({\bf \overline{16}})(-2)$ Weyl brane fermions
at the fixed point $(y_0,v_0)=(0,0)$ $\psi_{\bf \overline{120}}$,
$\psi_{\bf 16}^{(c)}$, and $\psi_{\bf \overline{16}}^{(d)}$
$(c,d=1,2,...,12)$. The matter content of the $SO(32)$ special GUT is
summarized in Table~\ref{tab:SO32-SU16-SO10-matter-content-6D}.
We will see what kind of roles each field has in detail in the
followings.

\begin{table}[t]
\begin{center}
\begin{tabular}{cccc}\hline
\rowcolor[gray]{0.8}
6D Bulk field&
$A_M$&$\Psi_{{\bf 32}+}^{(a)}$&$\Psi_{{\bf 32}-}^{(b)}$
\\\hline
$SO(32)$ &${\bf 496}$&${\bf 32}$&${\bf 32}$\\
$SO(5,1)$&${\bf 6}$&${\bf 4}_+$&${\bf 4}_-$\\
Orbifold BC&
&$\left(
\begin{array}{cc}
-&-\\
-&-\\
\end{array}
\right)$
&$\left(
\begin{array}{cc}
-&+\\
-&+\\
\end{array}
\right)$\\
\hline
\end{tabular}\\[0.5em]
\begin{tabular}{cccc}\hline
\rowcolor[gray]{0.8}
5D Brane field
&$\Phi_{\bf 86768}$&$\Phi_{\bf 496}$&$\Phi_{\bf 32}$\\
\hline
$SO(32)$ &${\bf 86768}$&${\bf 496}$&${\bf 32}$\\
$SO(4,1)$&{\bf 4}&{\bf 1}&{\bf 1}\\
Orbifold BC
&$\left(
\begin{array}{c}
+\\
-\\
\end{array}
\right)$
&$\left(
\begin{array}{c}
+\\
+\\
\end{array}
\right)$
&$\left(
\begin{array}{c}
+\\
+\\
\end{array}
\right)$\\
Spacetime  &$y=0$&$y=0$&$y=0$\\
\hline
\end{tabular}\\[0.5em]
\begin{tabular}{cccc}\hline
\rowcolor[gray]{0.8}
4D Brane field&$\psi_{\overline{\bf 120}}$&
$\psi_{{\bf 16}}^{(c)}$&$\psi_{\overline{\bf 16}}^{(d)}$\\
\hline
$SU(16)$&$\overline{\bf 120}$&${\bf 16}$&$\overline{\bf 16}$\\
$U(1)_Z$  &$0$&$0$&$-2$\\
$SL(2,\mathbb{C})$&$(1/2,0)$&$(1/2,0)$&$(1/2,0)$\\
Spacetime $(y,v)$ &$(0,0)$&$(0,0)$&$(0,0)$\\
\hline
\end{tabular}
\end{center}
\caption{The table shows the matter content in
the $SO(32)$ special GUT on $M^4\times T^2/\mathbb{Z}_2$.
The representations of $SO(32)$ and 6D, 5D, 4D Lorentz group, 
the orbifold BCs of 6D bulk fields and 5D brane fields, and 
the spacetime location of 5D and 4D brane fields
are shown.
\label{tab:SO32-SU16-SO10-matter-content-6D}}
\end{table}

First, a 6D $SO(32)$ bulk gauge boson $A_{M}$ is decomposed into 
a 4D gauge field $A_\mu$ and 5th and 6th dimensional gauge fields $A_y$
and $A_v$. The orbifold BCs of the 6D $SO(32)$ gauge field are given by
\begin{align}
\left(
\begin{array}{c}
A_\mu\\
A_y\\
A_v\\
\end{array}
\right)(x,y_j-y,v_j-v)
=P_{j{\bf 32}}
\left(
\begin{array}{c}
A_\mu\\
-A_y\\
-A_v\\
\end{array}
\right)(x,y_j+y,v_j+v)
P_{j{\bf 32}}^{-1},
\label{Eq:gauge-field-BCs}
\end{align}
where $P_{j{\bf 32}}$ is a projection matrix satisfying 
$(P_{j{\bf 32}})^2=I_{32}$. 
We consider the orbifold BCs $P_0$ and $P_1$ preserving $SO(32)$
symmetry, while the orbifold BCs $P_2$ and $P_3$ reduce $SO(32)$ to its
regular subgroup $SU(16)\times U(1)_Z$.
We take $P_{j{\bf 32}}$ as 
\begin{align}
P_{j{\bf 32}}=
\left\{
\begin{array}{ll}
\sigma_2\otimes I_2\otimes I_2\otimes I_2\otimes I_2&
\mbox{for}\ j=2,3\\
I_{32}\ &\mbox{for}\ j=0,1
\end{array}
\right..
\label{Eq:SO32-BCs-32}
\end{align}
In this case, the 4D $SO(32)$ ${\bf 496}$ gauge field $A_\mu$ have
Neumann BCs at the fixed points $(y_0,v_0)$ and 
$(y_1,v_1)$, while the 5th and 6th dimensional gauge
fields $A_y$ and $A_v$ have Dirichlet BCs because of the negative sign
in Eq.~(\ref{Eq:gauge-field-BCs}).
On the other hand, since $SO(32)$ symmetry is broken to 
$SU(16)\times U(1)_Z$ at the fixed points $(y_2,v_2)$ and $(y_3,v_3)$,
by using the branching rules of the $SO(32)$ adjoint representation 
${\bf 496}$ given in Eq.~(\ref{branching-rule-496}),
the $SU(16)\times U(1)_Z$ $\left(({\bf 255})(0)\oplus({\bf 1})(0)\right)$
and $\left(({\bf 120})(4)\oplus({\bf \overline{120}})(-4)\right)$ 
components of the 4D gauge field $A_\mu$ 
have Neumann and Dirichlet BCs at
the fixed points $(y_2,v_2)$ and $(y_3,v_3)$, respectively;
the $SU(16)\times U(1)_Z$ 
$\left(({\bf 255})(0)\oplus({\bf 1})(0)\right)$
and $\left(({\bf 120})(4)\oplus({\bf \overline{120}})(-4)\right)$ 
components of the 5th and 6th dimensional gauge 
fields $A_y$ and $A_v$ have Dirichlet and Neumann BCs, respectively.
Thus, since the $SU(16)\times U(1)_Z$ 
$\left(({\bf 255})(0)\oplus({\bf 1})(0)\right)$ components of the 4D
gauge field $A_\mu$ have four Neumann BCs at the four fixed points
$(y_j,v_j) (j=0,1,2,3)$, they have zero modes corresponding to 
4D $SU(16)$ and $U(1)_Z$ gauge fields;
since the other components of $A_\mu$ and any component of $A_y$ and
$A_v$ have four Dirichlet BCs or two Neumann and two Dirichlet BCs at the
four fixed points, they do not have zero modes.
The orbifold BCs reduce $SO(32)$ to $SU(16)\times U(1)_Z$. 
(Since there are no zero modes of the extra dimensional gauge field $A_y$
and $A_v$, we cannot rely on symmetry breaking known as the Hosotani
mechanism \cite{Hosotani:1983xw,Hosotani:1988bm} in this setup.)

To achieve the SM gauge symmetry $G_{\rm SM}$ at low energies, 
we consider the symmetry breaking sector via spontaneous
symmetry breaking. We introduce 
5D $SO(32)$ ${\bf 86768}$, ${\bf 496}$ and ${\bf 32}$ brane scalar fields 
$\Phi_{\bf 86768}$, $\Phi_{\bf 496}$ and $\Phi_{\bf 32}$ on the 5D brane
($y=0$). 
Their orbifold BCs are given by 
\begin{align}
\Phi_{\bf x}(x,v_\ell-v)=&
\eta_{\ell{\bf x}}P_{\ell{\bf x}}\Phi_{\bf x}(x,v_\ell+v),
\end{align}
where $\ell=0,2$, ${\bf x}$ stands for ${\bf 86768}$, ${\bf 496}$ and
${\bf 32}$, 
$\eta_{\ell{\bf x}}$ is a positive or negative sign, and
$P_{\ell{\bf x}}$ is a projection matrix. 
We take $\eta_{0{\bf 86768}}=-\eta_{2{\bf 86768}}=-1$ 
and $\eta_{\ell{\bf 496}}=\eta_{\ell{\bf 32}}=1$.
The tensor products of $SO(32)$ for ${\bf 496}$ and ${\bf 32}$ are given
in Ref.~\cite{Yamatsu:2015gut} by 
\begin{align}
{\bf 496}\otimes{\bf 496}=&
({\bf 86768})_{\rm S}
\oplus({\bf 35960})_{\rm S}
\oplus({\bf 527})_{\rm S}
\oplus({\bf 1})_{\rm S}
\oplus({\bf 122264})_{\rm A}
\oplus({\bf 496})_{\rm A},\\
{\bf 32}\otimes{\bf 32}=&
({\bf 527})_{\rm S}
\oplus({\bf 1})_{\rm S}
\oplus({\bf 496})_{\rm A}.
\end{align}
The branching rules of $SO(32)\supset SU(16)\times U(1)$ for 
${\bf 496}$ and ${\bf 86768}$ are given in
Eqs.~(\ref{branching-rule-496}) and (\ref{branching-rule-86768}),
respectively; 
for ${\bf 527}$, ${\bf 35960}$, and ${\bf 122264}$ are listed in
Ref.~\cite{Yamatsu:2015gut}.
For $\Phi_{\bf 86768}$, the $SU(16)\times U(1)$ 
$\left(({\bf 18240})(0)\oplus({\bf 14144})(0)\oplus({\bf 5440})(8)
\oplus({\bf \overline{5440}})(-8)\oplus({\bf 255})(0)\oplus({\bf 1})(0)
\right)$ components
have zero modes; for $\Phi_{\bf 496}$, the $SU(16)\times U(1)$ 
$\left(({\bf 255})(0)\oplus({\bf 1})(0)\right)$ components 
have zero modes;
and for $\Phi_{\bf 32}$, the $SU(16)\times U(1)$ $({\bf 16})(2)$
components have zero modes.
We assume that the nonvanishing VEV of the scalar field 
$\Phi_{\bf 86768}$ is responsible for breaking 
$(SO(32)\supset)SU(16)\times U(1)_Z$ to $SO(10)$; 
the nonvanishing VEV of the scalar field 
$\Phi_{\bf 32}$ breaks $(SO(32)\supset)SO(10)$ to $SU(5)$;
the nonvanishing VEV of $\Phi_{\bf 496}$ breaks $(SO(32)\supset)SU(5)$
to $G_{\rm SM}$.

The SM Weyl fermions are identified with zero modes of 6D $SO(32)$ 
${\bf 32}$ Weyl bulk fermions. The orbifold BCs of a 6D $SO(32)$ 
${\bf 32}$ positive or negative Weyl bulk fermion can be written by 
\begin{align}
\Psi_{{\bf 32}\pm}(x,y_j-y,v_j-v)&=
\eta_{j}(-i\Gamma^5\Gamma^6)
P_{j{\bf 32}}
\Psi_{{\bf 32}\pm}
(x,y_j+y,v_j+v),
\label{Eq:BC-SU(32)-fermion-32}
\end{align}
where the subscript of $\Psi$ $\pm$ stands for 6D chirality,
$\eta_{j}$ is a positive or negative sign,
$\prod_{j=0}^{3}\eta_j=1$,
$\Gamma^M$ $(M=1,2,\cdots,7)$ is a 6D gamma matrix, and 
$P_{j{\bf 32}}$ is give in Eq.~(\ref{Eq:SO32-BCs-32}).
Here, we check zero modes of, e.g., a 6D $SO(32)$ ${\bf 32}$ positive
Weyl fermion with orbifold BCs
$(\eta_{0},\eta_{1},\eta_{2},\eta_{3})=(-,-,-,-)$.
At fixed points $(y_0,v_0)$ and $(y_1,v_1)$, 
the 4D $SO(32)$ ${\bf 32}$ left-handed Weyl fermion components have
Neumann BCs, while the 4D $SO(32)$ ${\bf 32}$ right-handed Weyl fermion
components have Neumann BCs.
At fixed points $(y_0,v_0)$ and $(y_1,v_1)$, 
the 4D $SU(16)\times U(1)_Z$ $({\bf 16})(2)$ and 
$({\bf \overline{16}})(-2)$ left-handed Weyl fermion
components have Neumann and Dirichlet BCs, respectively,
while the 4D $SU(16)\times U(1)_Z$ $({\bf 16})(2)$ and 
$({\bf \overline{16}})(-2)$ right-handed Weyl fermion
components have Dirichlet and Neumann BCs, respectively.
In this case, only the 4D $SU(16)\times U(1)_Z$ $({\bf 16})(2)$ 
left-handed Weyl fermion has zero modes.
Note that only 6D $SO(32)$ ${\bf 32}$ positive Weyl fermions suffer from
6D gauge anomalies.

From the above, to realize three generations of the SM chiral fermions,
we introduce three sets of the pair of 6D $SO(32)$ ${\bf 32}$
positive and negative Weyl fermions to cancel out the 6D gauge
anomalies. More explicitly, each 
set of 6D Weyl fermions consists of a 6D $SO(32)$ ${\bf 32}$ positive
Weyl fermion with orbifold BCs
$(\eta_{0},\eta_{1},\eta_{2},\eta_{3})=(-,-,-,-)$ and 
a 6D negative one with orbifold BCs $(-,+,-,+)$.
Only the $SU(16)\times U(1)_Z$ $({\bf 16})(2)$ components of the
positive Weyl fermion have zero modes for its 4D left-handed Weyl
fermion components because its 4D left-handed Weyl fermion components
have Neumann BCs at all the fixed points.
The corresponding 4D right-handed Weyl fermion components have Dirichlet
BCs at all the fixed points.
The other components of the positive Weyl fermion and all the components
of the negative Weyl fermion have two Neumann and two Dirichlet BCs at
four fixed points $(y_j,v_j)$. 

Here, we check the contribution to 6D bulk and 4D brane anomalies from
the above 6D Weyl fermion sets. 
The fermion set does not contribute to 6D $SO(32)$ gauge anomaly because
of the same number of 6D $SO(32)$ ${\bf 32}$ positive and negative Weyl
fermions.
We need to check 4D gauge anomaly cancellation at four fixed points 
$(y_j,v_j) (j=0,1,2,3)$ by using 4D anomaly coefficients listed
in Ref.~\cite{Yamatsu:2015gut}.
At two fixed points $(y_j,v_j) (j=0,1)$, there is no 4D pure $SO(32)$
gauge anomaly because any 4D anomaly coefficient of $SO(32)$ is zero.
At the other two fixed points $(y_j,v_j) (j=2,3)$, there can be
4D pure $SU(16)$, pure $U(1)_Z$, mixed
$SU(16)-SU(16)-U(1)_Z$ and mixed $\mbox{grav.}-\mbox{grav.}-U(1)_Z$
anomalies.
At a fixed point $(y_3,v_3)$, the anomalies generated from the 6D
$SO(32)$ ${\bf 32}$ positive and negative Weyl fermions are canceled
each other; at the other fixed point $(y_2,v_2)$, the 6D $SO(32)$ 
${\bf 32}$ positive and negative Weyl fermions generate 
4D pure $SU(16)$, pure $U(1)_Z$, mixed
$SU(16)-SU(16)-U(1)_Z$ and mixed $\mbox{grav.}-\mbox{grav.}-U(1)_Z$
anomalies. We focus on how to cancel the 4D anomalies at the fixed point
$(y_2,v_2)$ below.

To achieve 4D gauge anomaly cancellation at the fixed point $(y_2,v_2)$,
we introduce 4D brane Weyl fermions in appropriate representations of
$SU(16)\times U(1)$.
First, we consider the pure $SU(16)$ gauge anomaly cancellation.
The 4D $SU(16)$ gauge anomaly of twelve 4D $SU(16)$ ${\bf 16}$
left-handed Weyl fermions is canceled out by 
the anomaly of a 4D $SU(16)$ ${\bf \overline{120}}$ Weyl fermion
\cite{Yamatsu:2017sgu}.
(It can be checked by using 4D $SU(16)$ anomaly coefficients listed in
Ref.~\cite{Yamatsu:2015gut}.)
When we introduce a 4D $SU(16)\times U(1)$ $({\bf \overline{120}})(0)$
Weyl fermion at $(y_2,v_2)$, its 4D $SU(16)$ anomaly cancels one
generated from the 6D bulk $SO(32)$ ${\bf 32}$ Weyl fermions.
Next, 4D pure $U(1)_Z$, mixed $SU(16)-SU(16)-U(1)_Z$, and mixed
$\mbox{grav.}-\mbox{grav.}-U(1)_Z$ anomalies can be canceled out by
introducing twelve 4D $SU(16)\times U(1)$ 
$\left(({\bf 16})(0)\oplus({\bf \overline{16}})(-2)\right)$ left-handed
Weyl fermions. This is because the matter content is vectorlike from the
view of the $U(1)_Z$ gauge theory. More explicitly, at the fixed point 
$(y_2,v_2)$, there are twelve $SU(16)\times U(1)_Z$ $({\bf 16})(2)$
left-handed Weyl fermions from the 6D $SO(32)$ ${\bf 32}$ Weyl fermions
and we introduced twelve $SU(16)\times U(1)_Z$ 
$({\bf \overline{16}})(-2)$ left-handed Weyl brane fermions.
Also, the 4D $SU(16)\times U(1)$ 
$\left(({\bf 16})(0)\oplus({\bf \overline{16}})(-2)\right)$ Weyl
fermions do not generate 4D pure $SU(16)$ anomaly.
Note that the above brane fermions become vectorlike when 
$SU(16)\times U(1)$ symmetry is broken to $SO(10)$, so there is no
exotic chiral fermions at low energies, where 
the branching rules of $SU(16)\supset SO(10)$ for an $SU(16)$ complex
representation ${\bf \overline{120}}$ (${\bf 120}$) is identified with 
an $SO(10)$ real representation ${\bf 120}$: 
\begin{align}
{\bf \overline{120}}={\bf 120}\ \ \
({\bf 120}={\bf 120}).
\end{align}

\section{Summary and discussion}
\label{Sec:Summary-discussion}

In this paper, we constructed an $SO(32)$ special GUT by using a
special breaking $SU(16)$ to $SO(10)$. In this framework, the zero modes
of the 6D $SO(32)$ ${\bf 32}$ Weyl fermion can be identified with
one generation of quarks and leptons;
the 6D $SO(32)$ and the 4D $SU(16)\times U(1)$ gauge anomalies on the
fixed points allow a three generation model of quarks and leptons in 6D
framework; as in the $SU(16)$ special GUT \cite{Yamatsu:2017sgu}, 
exotic chiral fermions do not exist due to a special feature of
the $SU(16)$ complex representation ${\bf \overline{120}}$ once we take
into account the symmetry breaking of $SO(32)$ to $SO(10)$.

In this paper, we simply assumed that the nonvanishing VEV of a scalar
field $\Phi_{\bf 86768}$ breaks $(SO(32)\supset)SU(16)\times U(1)_Z$ to
$SO(10)$. 
Instead, we may consider dynamical symmetry breaking scenario 
\cite{Raby:1979my,Dimopoulos:1979es,Farhi:1980xs,Peskin:1980gc,Miransky:1988xi,Miransky:1989ds,Bardeen:1989ds,Kugo:1994qr,Nambu:1961tp}
to realize the special breaking $SU(16)$ to $SO(10)$.
Its breaking can be realized by using the pair condensation of a fermion
in the $SO(32)$ adjoint representation ${\bf 496}$ or the $SU(16)$
second-rank anti-symmetric tensor ${\bf 120}$ $({\bf \overline{120}})$.
The analysis will be reported in a separate paper
\cite{Kugo:2017dsb}. 
(The dynamical symmetry breaking of $SU(16)$ to its special subgroup
$SO(10)$ is essentially the same as one of $E_6$ to its special
subgroups $F_4$ and $USp(8)$ or $G_2$ or $SU(3)$ discussed in
Ref.~\cite{Kugo:1994qr}.) 

To cancel 4D pure $SU(16)$, pure $U(1)_Z$, mixed $SU(16)-SU(16)-U(1)_Z$
and mixed $\mbox{grav.}-\mbox{grav.}-U(1)_Z$ anomalies on a fixed point,
we introduced several brane Weyl fermions. For the mixed anomalies, one
may rely on Green-Schwarz (GS) anomaly cancellation mechanism
\cite{Green1984} for 4D version \cite{Binetruy:1996uv,Kojima:2017qbt} by
introducing a pseudo-scalar field that transforms non-linearly under the
anomalous $U(1)$ symmetry.

\section*{Acknowledgments}

The author would like to thank Yutaka Hosotani, Kentaro Kojima,
Taichiro Kugo, Shogo Kuwakino, Kenji Nishiwaki, and Shohei Uemura 
for valuable comments. 

\bibliographystyle{utphys} 
\bibliography{../../arxiv/reference}

\providecommand{\href}[2]{#2}\begingroup\raggedright\begin{thebibliography}{10}

\bibitem{Georgi:1974sy}
H.~Georgi and S.~L. Glashow, ``{Unity of All Elementary Particle Forces},''
\href{http://dx.doi.org/10.1103/PhysRevLett.32.438}{{\em Phys. Rev. Lett.}
  {\bfseries 32} (1974) 438--441}.

\bibitem{Slansky:1981yr}
R.~Slansky, ``{Group Theory for Unified Model Building},''
\href{http://dx.doi.org/10.1016/0370-1573(81)90092-2}{{\em Phys. Rept.}
  {\bfseries 79} (1981) 1--128}.

\bibitem{Yamatsu:2015gut}
N.~Yamatsu, ``{Finite-Dimensional Lie Algebras and Their Representations for
  Unified Model Building},''
\href{http://arxiv.org/abs/1511.08771}{{\ttfamily arXiv:1511.08771 [hep-ph]}}.

\bibitem{Inoue:1977qd}
K.~Inoue, A.~Kakuto, and Y.~Nakano, ``{Unification of the Lepton-Quark World by
  the Gauge Group SU(6)},''
\href{http://dx.doi.org/10.1143/PTP.58.630}{{\em Prog.Theor.Phys.} {\bfseries
  58} (1977) 630}.

\bibitem{Fritzsch:1974nn}
H.~Fritzsch and P.~Minkowski, ``{Unified Interactions of Leptons and
  Hadrons},''
\href{http://dx.doi.org/10.1016/0003-4916(75)90211-0}{{\em Ann. Phys.}
  {\bfseries 93} (1975) 193--266}.

\bibitem{Ida:1980ea}
M.~Ida, Y.~Kayama, and T.~Kitazoe, ``{Inclusion of Generations in SO(14)},''
\href{http://dx.doi.org/10.1143/PTP.64.1745}{{\em Prog. Theor. Phys.}
  {\bfseries 64} (1980) 1745}.

\bibitem{Fujimoto:1981bv}
Y.~Fujimoto, ``{SO(18) Unification},''
\href{http://dx.doi.org/10.1103/PhysRevD.26.3183}{{\em Phys. Rev.} {\bfseries
  D26} (1982) 3183}.

\bibitem{Gursey:1975ki}
F.~Gursey, P.~Ramond, and P.~Sikivie, ``{A Universal Gauge Theory Model Based
  on $E_6$},''
\href{http://dx.doi.org/10.1016/0370-2693(76)90417-2}{{\em Phys. Lett.}
  {\bfseries B60} (1976) 177}.

\bibitem{Kawamura:2013rj}
Y.~Kawamura and T.~Miura, ``{Classification of Standard Model Particles in
  $E_6$ Orbifold Grand Unified Theories},''
  \href{http://dx.doi.org/10.1142/S0217751X13500553}{{\em Int. J. Mod. Phys.}
  {\bfseries A28} (2013) 1350055},
\href{http://arxiv.org/abs/1301.7469}{{\ttfamily arXiv:1301.7469 [hep-ph]}}.

\bibitem{Kojima:2017qbt}
K.~Kojima, K.~Takenaga, and T.~Yamashita, ``{The Standard Model Gauge Symmetry
  from Higher-Rank Unified Groups in Grand Gauge-Higgs Unification Models},''
  \href{http://dx.doi.org/10.1007/JHEP06(2017)018}{{\em JHEP} {\bfseries 06}
  (2017) 018},
\href{http://arxiv.org/abs/1704.04840}{{\ttfamily arXiv:1704.04840 [hep-ph]}}.

\bibitem{Kojima:2011ad}
K.~Kojima, K.~Takenaga, and T.~Yamashita, ``{Grand Gauge-Higgs Unification},''
  \href{http://dx.doi.org/10.1103/PhysRevD.84.051701}{{\em Phys. Rev.}
  {\bfseries D84} (2011) 051701},
\href{http://arxiv.org/abs/1103.1234}{{\ttfamily arXiv:1103.1234 [hep-ph]}}.

\bibitem{Kojima:2016fvv}
K.~Kojima, K.~Takenaga, and T.~Yamashita, ``{Gauge Symmetry Breaking Patterns
  in an SU(5) Grand Gauge-Higgs Unification Model},''
  \href{http://dx.doi.org/10.1103/PhysRevD.95.015021}{{\em Phys. Rev.}
  {\bfseries D95} no.~1, (2017) 015021},
\href{http://arxiv.org/abs/1608.05496}{{\ttfamily arXiv:1608.05496 [hep-ph]}}.

\bibitem{Burdman:2002se}
G.~Burdman and Y.~Nomura, ``{Unification of Higgs and Gauge Fields in
  Five-Dimensions},''
  \href{http://dx.doi.org/10.1016/S0550-3213(03)00088-9}{{\em Nucl. Phys.}
  {\bfseries B656} (2003) 3--22},
\href{http://arxiv.org/abs/hep-ph/0210257}{{\ttfamily arXiv:hep-ph/0210257
  [hep-ph]}}.

\bibitem{Lim:2007jv}
C.~Lim and N.~Maru, ``{Towards a Realistic Grand Gauge-Higgs Unification},''
  \href{http://dx.doi.org/10.1016/j.physletb.2007.07.053}{{\em Phys.Lett.}
  {\bfseries B653} (2007) 320--324},
\href{http://arxiv.org/abs/0706.1397}{{\ttfamily arXiv:0706.1397 [hep-ph]}}.

\bibitem{Kim:2002im}
H.~D. Kim and S.~Raby, ``{Unification in 5-D SO(10)},''
  \href{http://dx.doi.org/10.1088/1126-6708/2003/01/056}{{\em JHEP} {\bfseries
  01} (2003) 056},
\href{http://arxiv.org/abs/hep-ph/0212348}{{\ttfamily arXiv:hep-ph/0212348
  [hep-ph]}}.

\bibitem{Fukuyama:2008pw}
T.~Fukuyama and N.~Okada, ``{A Simple SO(10) GUT in Five Dimensions},''
  \href{http://dx.doi.org/10.1103/PhysRevD.78.015005}{{\em Phys. Rev.}
  {\bfseries D78} (2008) 015005},
\href{http://arxiv.org/abs/0803.1758}{{\ttfamily arXiv:0803.1758 [hep-ph]}}.

\bibitem{Hosotani:2015hoa}
Y.~Hosotani and N.~Yamatsu, ``{Gauge-Higgs Grand Unification},''
  \href{http://dx.doi.org/10.1093/ptep/ptv153}{{\em Prog. Theor. Exp. Phys.}
  {\bfseries 2015} (2015) 111B01},
\href{http://arxiv.org/abs/1504.03817}{{\ttfamily arXiv:1504.03817 [hep-ph]}}.

\bibitem{Hosotani:2015wmb}
Y.~Hosotani and N.~Yamatsu, ``{Gauge-Higgs Grand Unification},'' {\em PoS}
  {\bfseries PLANCK2015} (2015) 058,
\href{http://arxiv.org/abs/1511.01674}{{\ttfamily arXiv:1511.01674 [hep-ph]}}.

\bibitem{Yamatsu:2015rge}
N.~Yamatsu, ``{Gauge Coupling Unification in Gauge-Higgs Grand Unification},''
  \href{http://dx.doi.org/10.1093/ptep/ptw023}{{\em Prog. Theor. Exp. Phys.}
  {\bfseries 2016} (2016) 043B02},
\href{http://arxiv.org/abs/1512.05559}{{\ttfamily arXiv:1512.05559 [hep-ph]}}.

\bibitem{Furui:2016owe}
A.~Furui, Y.~Hosotani, and N.~Yamatsu, ``{Toward Realistic Gauge-Higgs Grand
  Unification},'' \href{http://dx.doi.org/10.1093/ptep/ptw116}{{\em Prog.
  Theor. Exp. Phys.} {\bfseries 2016} (2016) 093B01},
\href{http://arxiv.org/abs/1606.07222}{{\ttfamily arXiv:1606.07222 [hep-ph]}}.

\bibitem{Hosotani:2016njs}
Y.~Hosotani, ``{Gauge-Higgs EW and Grand Unification},''
  \href{http://dx.doi.org/10.1142/S0217751X16300313}{{\em Int. J. Mod. Phys.}
  {\bfseries A31} no.~20n21, (2016) 1630031},
\href{http://arxiv.org/abs/1606.08108}{{\ttfamily arXiv:1606.08108 [hep-ph]}}.

\bibitem{Hosotani:2017ghg}
Y.~Hosotani and N.~Yamatsu, ``{Gauge-Higgs Seesaw Mechanism in Six-Dimensional
  Grand Unification},''
\href{http://arxiv.org/abs/1706.03503}{{\ttfamily arXiv:1706.03503 [hep-ph]}}.

\bibitem{Yamatsu:2017sgu}
N.~Yamatsu, ``{Special Grand Unification},''
  \href{http://dx.doi.org/10.1093/ptep/ptx088}{{\em Prog. Theor. Exp. Phys.}
  {\bfseries 2017} (2017) 061B01},
\href{http://arxiv.org/abs/1704.08827}{{\ttfamily arXiv:1704.08827 [hep-ph]}}.

\bibitem{Dynkin:1957ek}
E.~Dynkin, ``{Maximal Subgroups of the Classical Groups},'' {\em Amer. Math.
  Soc. Transl.} {\bfseries 6} (1957) 245.

\bibitem{Dynkin:1957um}
E.~Dynkin, ``{Semisimple Subalgebras of Semisimple Lie Algebras},''
{\em Amer. Math. Soc. Transl.} {\bfseries 6} (1957) 111.

\bibitem{Cahn:1985wk}
R.~Cahn, {\em {Semi-Simple Lie Algebras and Their Representations}}.
\newblock Benjamin-Cummings Publishing Company,
1985.
\newblock

\bibitem{Polchinski1998a}
J.~Polchinski, {\em {String Theory I -An Introduction to Bosonic String-}}.
\newblock Cambridge University Press, 1998.

\bibitem{Polchinski1998b}
J.~Polchinski, {\em {String Theory II -Superstring Theory and Beyond}}.
\newblock Cambridge University Press, 1998.

\bibitem{Green1984}
M.~B. Green and J.~H. Schwarz, ``{Anomaly Cancellation in Supersymmetric D=10
  Gauge Theory and Superstring Theory},''
\href{http://dx.doi.org/10.1016/0370-2693(84)91565-X}{{\em Phys. Lett.}
  {\bfseries B149} (1984) 117--122}.

\bibitem{Green:1984ed}
M.~B. Green and J.~H. Schwarz, ``{Infinity Cancellations in SO(32) Superstring
  Theory},''
\href{http://dx.doi.org/10.1016/0370-2693(85)90816-0}{{\em Phys. Lett.}
  {\bfseries 151B} (1985) 21--25}.

\bibitem{Derendinger:1985kk}
J.~P. Derendinger, L.~E. Ibanez, and H.~P. Nilles, ``{On the Low-Energy d = 4,
  N=1 Supergravity Theory Extracted from the d = 10, N=1 Superstring},''
\href{http://dx.doi.org/10.1016/0370-2693(85)91033-0}{{\em Phys. Lett.}
  {\bfseries 155B} (1985) 65--70}.

\bibitem{Giedt:2003an}
J.~Giedt, ``{Z(3) Orbifolds of the SO(32) Heterotic String. 1. Wilson Line
  Embeddings},'' \href{http://dx.doi.org/10.1016/j.nuclphysb.2003.08.031}{{\em
  Nucl. Phys.} {\bfseries B671} (2003) 133--147},
\href{http://arxiv.org/abs/hep-th/0301232}{{\ttfamily arXiv:hep-th/0301232
  [hep-th]}}.

\bibitem{Choi:2004wn}
K.-S. Choi, S.~Groot~Nibbelink, and M.~Trapletti, ``{Heterotic SO(32) Model
  Building in Four Dimensions},''
  \href{http://dx.doi.org/10.1088/1126-6708/2004/12/063}{{\em JHEP} {\bfseries
  12} (2004) 063},
\href{http://arxiv.org/abs/hep-th/0410232}{{\ttfamily arXiv:hep-th/0410232
  [hep-th]}}.

\bibitem{Blumenhagen:2005pm}
R.~Blumenhagen, G.~Honecker, and T.~Weigand, ``{Supersymmetric (Non-)Abelian
  Bundles in the Type I and SO(32) Heterotic String},''
  \href{http://dx.doi.org/10.1088/1126-6708/2005/08/009}{{\em JHEP} {\bfseries
  08} (2005) 009},
\href{http://arxiv.org/abs/hep-th/0507041}{{\ttfamily arXiv:hep-th/0507041
  [hep-th]}}.

\bibitem{Nilles:2006np}
H.~P. Nilles, S.~Ramos-Sanchez, P.~K.~S. Vaudrevange, and A.~Wingerter,
  ``{Exploring the SO(32) Heterotic String},''
  \href{http://dx.doi.org/10.1088/1126-6708/2006/04/050}{{\em JHEP} {\bfseries
  04} (2006) 050},
\href{http://arxiv.org/abs/hep-th/0603086}{{\ttfamily arXiv:hep-th/0603086
  [hep-th]}}.

\bibitem{Ito:2010df}
M.~Ito, S.~Kuwakino, N.~Maekawa, S.~Moriyama, K.~Takahashi, K.~Takei,
  S.~Teraguchi, and T.~Yamashita, ``{$E_6$ Grand Unified Theory with Three
  Generations from Heterotic String},''
  \href{http://dx.doi.org/10.1103/PhysRevD.83.091703}{{\em Phys. Rev.}
  {\bfseries D83} (2011) 091703},
\href{http://arxiv.org/abs/1012.1690}{{\ttfamily arXiv:1012.1690 [hep-ph]}}.

\bibitem{Ito:2011ng}
M.~Ito, S.~Kuwakino, N.~Maekawa, S.~Moriyama, K.~Takahashi, K.~Takei,
  S.~Teraguchi, and T.~Yamashita, ``{Heterotic $E_6$ GUTs and Partition
  Functions},'' \href{http://dx.doi.org/10.1007/JHEP12(2011)100}{{\em JHEP}
  {\bfseries 12} (2011) 100},
\href{http://arxiv.org/abs/1104.0765}{{\ttfamily arXiv:1104.0765 [hep-th]}}.

\bibitem{Abe:2015mua}
H.~Abe, T.~Kobayashi, H.~Otsuka, and Y.~Takano, ``{Realistic Three-Generation
  Models from SO(32) Heterotic String Theory},''
  \href{http://dx.doi.org/10.1007/JHEP09(2015)056}{{\em JHEP} {\bfseries 09}
  (2015) 056},
\href{http://arxiv.org/abs/1503.06770}{{\ttfamily arXiv:1503.06770 [hep-th]}}.

\bibitem{Abe:2015xua}
H.~Abe, T.~Kobayashi, H.~Otsuka, Y.~Takano, and T.~H. Tatsuishi, ``{Gauge
  Coupling Unification in SO(32) Heterotic String Theory with Magnetic
  Fluxes},'' \href{http://dx.doi.org/10.1093/ptep/ptw038}{{\em PTEP} {\bfseries
  2016} no.~5, (2016) 053B01},
\href{http://arxiv.org/abs/1507.04127}{{\ttfamily arXiv:1507.04127 [hep-ph]}}.

\bibitem{Abe:2016eyh}
H.~Abe, T.~Kobayashi, H.~Otsuka, Y.~Takano, and T.~H. Tatsuishi, ``{Flavor
  Structure in $SO(32)$ Heterotic String Theory},''
  \href{http://dx.doi.org/10.1103/PhysRevD.94.126020}{{\em Phys. Rev.}
  {\bfseries D94} no.~12, (2016) 126020},
\href{http://arxiv.org/abs/1605.00898}{{\ttfamily arXiv:1605.00898 [hep-ph]}}.

\bibitem{Kawamura:1999nj}
Y.~Kawamura, ``{Gauge Symmetry Breaking from Extra Space $S^1/Z_2$},''
  \href{http://dx.doi.org/10.1143/PTP.103.613}{{\em Prog. Theor. Phys.}
  {\bfseries 103} (2000) 613--619},
\href{http://arxiv.org/abs/hep-ph/9902423}{{\ttfamily arXiv:hep-ph/9902423
  [hep-ph]}}.

\bibitem{Kawamura:2000ev}
Y.~Kawamura, ``{Triplet-Doublet Splitting, Proton Stability and Extra
  Dimension},'' \href{http://dx.doi.org/10.1143/PTP.105.999}{{\em Prog. Theor.
  Phys.} {\bfseries 105} (2001) 999--1006},
\href{http://arxiv.org/abs/hep-ph/0012125}{{\ttfamily arXiv:hep-ph/0012125}}.

\bibitem{Higgs:1964ia}
P.~W. Higgs, ``{Broken Symmetries, Massless Particles and Gauge Fields},''
\href{http://dx.doi.org/10.1016/0031-9163(64)91136-9}{{\em Phys. Lett.}
  {\bfseries 12} (1964) 132--133}.

\bibitem{Higgs:1964pj}
P.~W. Higgs, ``{Broken Symmetries and the Masses of Gauge Bosons},''
\href{http://dx.doi.org/10.1103/PhysRevLett.13.508}{{\em Phys. Rev. Lett.}
  {\bfseries 13} (1964) 508--509}.

\bibitem{Li:1973mq}
L.-F. Li, ``{Group Theory of the Spontaneously Broken Gauge Symmetries},''
\href{http://dx.doi.org/10.1103/PhysRevD.9.1723}{{\em Phys. Rev.} {\bfseries
  D9} (1974) 1723--1739}.

\bibitem{Meljanac:1982rc}
S.~Meljanac, M.~Milosevic, and S.~Pallua, ``{Extrema of Higgs Potential and
  Higher Representations},''
\href{http://dx.doi.org/10.1103/PhysRevD.26.2936}{{\em Phys. Rev.} {\bfseries
  D26} (1982) 2936--2939}.

\bibitem{Randall:1999ee}
L.~Randall and R.~Sundrum, ``{A Large Mass Hierarchy from a Small Extra
  Dimension},'' \href{http://dx.doi.org/10.1103/PhysRevLett.83.3370}{{\em Phys.
  Rev. Lett.} {\bfseries 83} (1999) 3370--3373},
\href{http://arxiv.org/abs/hep-ph/9905221}{{\ttfamily arXiv:hep-ph/9905221}}.

\bibitem{Hosotani:1983xw}
Y.~Hosotani, ``{Dynamical Mass Generation by Compact Extra Dimensions},''
\href{http://dx.doi.org/10.1016/0370-2693(83)90170-3}{{\em Phys.Lett.}
  {\bfseries B126} (1983) 309}.

\bibitem{Hosotani:1988bm}
Y.~Hosotani, ``{Dynamics of Nonintegrable Phases and Gauge Symmetry
  Breaking},''
\href{http://dx.doi.org/10.1016/0003-4916(89)90015-8}{{\em Annals Phys.}
  {\bfseries 190} (1989) 233}.

\bibitem{Raby:1979my}
S.~Raby, S.~Dimopoulos, and L.~Susskind, ``{Tumbling Gauge Theories},''
\href{http://dx.doi.org/10.1016/0550-3213(80)90093-0}{{\em Nucl.Phys.}
  {\bfseries B169} (1980) 373}.

\bibitem{Dimopoulos:1979es}
S.~Dimopoulos and L.~Susskind, ``{Mass Without Scalars},''
  \href{http://dx.doi.org/10.1016/0550-3213(79)90364-X}{{\em Nucl. Phys.}
  {\bfseries B155} (1979) 237--252}.
[2,930(1979)].

\bibitem{Farhi:1980xs}
E.~Farhi and L.~Susskind, ``{Technicolor},''
\href{http://dx.doi.org/10.1016/0370-1573(81)90173-3}{{\em Phys. Rept.}
  {\bfseries 74} (1981) 277}.

\bibitem{Peskin:1980gc}
M.~E. Peskin, ``{The Alignment of the Vacuum in Theories of Technicolor},''
\href{http://dx.doi.org/10.1016/0550-3213(80)90051-6}{{\em Nucl.Phys.}
  {\bfseries B175} (1980) 197--233}.

\bibitem{Miransky:1988xi}
V.~A. Miransky, M.~Tanabashi, and K.~Yamawaki, ``{Dynamical Electroweak
  Symmetry Breaking with Large Anomalous Dimension and t Quark Condensate},''
\href{http://dx.doi.org/10.1016/0370-2693(89)91494-9}{{\em Phys. Lett.}
  {\bfseries B221} (1989) 177--183}.

\bibitem{Miransky:1989ds}
V.~A. Miransky, M.~Tanabashi, and K.~Yamawaki, ``{Is the t Quark Responsible
  for the Mass of W and Z Bosons?},''
\href{http://dx.doi.org/10.1142/S0217732389001210}{{\em Mod. Phys. Lett.}
  {\bfseries A4} (1989) 1043}.

\bibitem{Bardeen:1989ds}
W.~A. Bardeen, C.~T. Hill, and M.~Lindner, ``{Minimal Dynamical Symmetry
  Breaking of the Standard Model},''
\href{http://dx.doi.org/10.1103/PhysRevD.41.1647}{{\em Phys. Rev.} {\bfseries
  D41} (1990) 1647}.

\bibitem{Kugo:1994qr}
T.~Kugo and J.~Sato, ``{Dynamical Symmetry Breaking in an $E_6$ GUT Model},''
  \href{http://dx.doi.org/10.1143/ptp/91.6.1217, 10.1143/PTP.91.1217}{{\em
  Prog. Theor. Phys.} {\bfseries 91} (1994) 1217--1238},
\href{http://arxiv.org/abs/hep-ph/9402357}{{\ttfamily arXiv:hep-ph/9402357
  [hep-ph]}}.

\bibitem{Nambu:1961tp}
Y.~Nambu and G.~Jona-Lasinio, ``{Dynamical Model of Elementary Particles Based
  on an Analogy with Superconductivity. I},''
\href{http://dx.doi.org/10.1103/PhysRev.122.345}{{\em Phys. Rev.} {\bfseries
  122} (1961) 345--358}.

\bibitem{Kugo:2017dsb}
T.~Kugo and N.~Yamatsu. in preparation.

\bibitem{Binetruy:1996uv}
P.~Binetruy and E.~Dudas, ``{Gaugino Condensation and the Anomalous U(1)},''
  \href{http://dx.doi.org/10.1016/S0370-2693(96)01305-6}{{\em Phys. Lett.}
  {\bfseries B389} (1996) 503--509},
\href{http://arxiv.org/abs/hep-th/9607172}{{\ttfamily arXiv:hep-th/9607172
  [hep-th]}}.

\end{thebibliography}\endgroup

\end{document}